# Dynamic response of the dielectric and electro-optic properties of epitaxial ferroelectric thin films

B. H. Hoerman, B.M.Nichols, and B. W. Wessels

Department of Materials Science and Engineering and
Materials Research Center, Northwestern University, Evanston, IL 60208

April 24, 2002

**Abstract**

An analysis of the dynamic dielectric and electro-optic relaxation response of thin film ferroelectrics is presented. The analysis is based upon the relaxation of ferroelectric domains with a continuous distribution of sizes given by percolation theory. The resulting temporal response is described by the expression $\Phi(t) \propto t^{-m} \exp(-(t/\tau)^{\beta})$. The analysis was applied to $KNbO_3$ thin films. Measurements of the polarization, birefringence and dielectric transients show qualitative agreement with the model over 11 orders of magnitude in time.

## I. INTRODUCTION

The dynamic response of the dielectric and non-linear optical properties of ferroelectrics is of fundamental interest for elucidating the microscopic mechanisms of the ferroelectric (FE) state. Both of these properties are influenced by the domain structure, which is in turn dependent on the microstructure. In thin film FEs the second harmonic generation (SHG)[1] and electro-optic (EO)[2] coefficients are typically an order of magnitude lower than in bulk materials. The reduction is attributed to the presence of a complex domain structure.[1,2] Upon application of an electric field these domains align, leading to an increase in the non-linear optical susceptibilities.[1-4] This poling process has been observed to occur on time scales ranging from 1 to 1000 s for SHG,[1] spontaneous birefringence,[5] and the EO effect.[2] Moreover, the alignment process in thin films, in contrast to single crystal FEs, has been observed to be completely reversible[1] upon removal of the poling field.[2-4] The competition of electrostatic and elastic interactions between domains is believed to be the source of the reversibility of the poling process.[1,4,6] Recently we have examined the influence of domain dynamics on the transient response of the non-linear optical properties of FE thin films on time scales from 6 ns to 100 ms.[3,7] The transient response of the EO effect follows a power law, $\sim t^{-m}$. This dependence was tentatively attributed to the relaxation of ferroelectric domains.

A similar power law time response has been observed for the dielectric properties of thin film FEs.[8,9] The time response of the relaxation current responsible for the dielectric loss has been observed to follow a power law time response, $t^{-m}$, or frequency response, $\omega^{m-1}$, with $0 < m < 1$. The power law dependence of the relaxation current has been observed to extend over 12 orders of magnitude in frequency.[9] The physical origin of the power law relaxation in thin film FEs is not understood, although processes involving FE domains are believed to play an important role.[10]

To explain the dielectric relaxation or switching current of ferroelectrics, a model based on the nucleation and growth of FE domains under an applied electric field has been previously developed.[11] The model assumes: (1) a random distribution of pre-existing, aligned domains with a uniform radius, $R_C$; (2) a constant nucleation rate of aligned domains with a radius $R_C$; (3) the growth of the domain boundary proceeds at a constant velocity and can be categorized as 1-, 2-, or 3-dimensional; and (4) no cooperative interaction between domains. The model predicts a stretched exponential growth of the volume fraction of the aligned domains, $1 - \exp(-(t/\tau)^\beta)$, where $t$ is the time, and $\beta$ and $\tau$ are constants. The value of $\beta$ depends on the dimensionality of the domain growth and can take on integer values of 1, 2, or 3. The expression describes experimental switching current data in many cases, although non-integer values of $\beta$ are often required.[11,12] For $t \ll \tau$, the stretched exponential reduces to a power law, $(t/\tau)^\beta$. However, the theory cannot account for the fractional values of $\beta$ typically observed.[11,12] Thus to describe such cases a different explanation of the power law response must be found.

While the previous model assumes pre-existing domains of uniform radius, the domain structure in thin films is very different. In contrast to the large sizes and regular shapes of bulk FE domains, irregularly shaped, nano-scale domains with a distribution of sizes have been observed in thin films.[1,10,13,14] The EO response of these domains to microwaves (2–4 GHz) has been studied.[10,14] Measurements indicate the switching process of the nanodomains take place over very short time scales.[12]

To fill in the gap between experiment and existing theory we have developed a model for the dynamic dielectric and EO response in thin film FEs that takes into account the distribution of domain sizes observed in thin film systems. The model predicts a temporal response that is described by the expression $\Phi(t) \propto t^{-m} \exp(-(t/\tau)^{\beta})$. The model is compared to transient measurements of the EO coefficient, the polarization, and the dielectric constant of epitaxial $KNbO_3$ thin films. These measurements support the proposed model in that the temporal response of all three properties agrees with the predicted response over time scales ranging from $\leq 6$ ns to $\sim 10^3$ s.

## II. THEORETICAL ANALYSIS

The proposed model describes the dynamic response in terms of a distribution of relaxation times,[15] in contrast to a "normal" FE system where the response is described by a single relaxation time. Since the relaxation rate depends on domain size, the broadened response occurs as a result of the distribution of domain sizes observed in the thin films.[16]

The relaxation function, $\Phi(t)$, for the system consisting of $\Omega$ dynamically correlated domains is assumed to be of the form:[15]

$$\Phi(t) = \sum_{i=0}^{\Omega} \phi(V_i) \exp\left(-\frac{t}{\tau_i}\right) \qquad (1)$$

where $t$ is the time, and $\phi(V_i)$ is the magnitude of the response for a domain with volume $V_i$ and relaxation time $\tau_i$. For FE materials, it has been previously proposed that the relaxation time, $\tau_i$, is directly proportional to the domain volume, $V_i$:[10,12,14,16-18]

$$\tau_i(V_i) = \frac{V_i}{B} \qquad (2)$$

where $B$ is a constant.[15] This expression accounts for the coupling between FE domains through electrostatic and mechanical interactions.[19]

For the EO effect the response, $\phi(V_i)$, is proportional to the cross-sectional area of the domain multiplied by the path length,[1,2] or equivalently the volume. Hence:

$$\Phi(t) = \sum_{i=0}^{\Omega} \phi' \, V_i \, \exp\left(-B\frac{t}{V_i}\right) \tag{3}$$

where the constant $\phi'$ represents the response of each domain per unit volume, ie. the spontaneous birefringence, polarization, or EO coefficient.

For a continuous distribution of domain sizes, the summation may be approximated by an integral by introducing a domain volume distribution, $N(V)$. It is supposed[20] that the form of $N(V)$ is given by percolation theory, which is often used to characterize quenched systems with local disorder[21,22] and has been used to model the relaxation of dielectrics.[15]

For a percolative distribution of FE domain volumes:[21]

$$N(V) = N_o \, V^{-\gamma} \tag{4}$$

where $N_o$ is a normalization constant and $\gamma$ is a critical exponent related to the fractal dimension, $D_F$, of the FE domains given by:[21]

$$D_F = \frac{D_E}{\gamma - 1} \tag{5}$$

where $D_E$ is the Euclidean dimension of the embedding space, $D_E = 3$. Physically $D_F$ can take on values spanning the Euclidean dimensions from 3 to 2.[21] This places limits on $\gamma$ ranging from 2 to 2.5, respectively. For percolation in three dimensions ($D_E = 3$), expected values are $D_F \approx 2.525 \pm$

0.01 and $\gamma \approx 2.19$. Recent reports of the fractal properties of FE domains support these values.[23-25]

The distribution given in Eq. (4) is restricted to a finite range of domain sizes.[21] For a continuous distribution of domain sizes between the limits $V_{max}$ (the volume of the largest domain) and $V_{min}$ (the volume of the smallest domains) the relaxation function is given by:

$$\Phi(t) = \phi' \int_{V_{min}}^{V_{max}} N(V) V \exp\left(-B\frac{t}{V}\right) dV \tag{6}$$

Upon substitution of Eq. (4) for $N(V)$,

$$\Phi(t) = \phi' N_o \int_{V_{min}}^{V_{max}} V^{1-\gamma} \exp\left(-B\frac{t}{V}\right) dV \tag{7}$$

Solving the integral:[26]

$$\Phi(t) = \phi' N_o (B t)^{2-\gamma} \left[\Gamma(\gamma-2, \tfrac{tB}{V_{max}}) - \Gamma(\gamma-2, \tfrac{tB}{V_{min}})\right] \tag{8}$$

where $\Gamma(x, y)$ is the incomplete gamma function defined as:[26]

$$\Gamma(x, y) = \int_{y}^{\infty} z^{x-1} e^{-z} dz \tag{9}$$

In the long-time limit, $t \gg V_{min}/B$, the response in Eq. (8) becomes:

$$\Phi(t) = \phi' N_o (B t)^{2-\gamma} \Gamma(\gamma-2, \tfrac{tB}{V_{max}}) \tag{10}$$

For a constant $0 < \gamma-2 < 0.5$, the function $\Gamma(\gamma-2, t B/V_{max})$ is well approximated by a stretched exponential with $0 < \beta < 1$, and $\Phi(t)$ is approximately:

$$\Phi(t) \propto t^{2-\gamma} \exp\left[-\left(\frac{t B}{V_{max}}\right)^{\beta}\right] = t^{2-\gamma} \exp-(t/\tau)^{\beta} \tag{11}$$

For the short-time limit $t \ll V_{max}/B$ (the relaxation time of the largest domains), the relaxation is given by:

$$\Phi(t) \propto t^{2-\gamma} = t^{-m} \qquad (12)$$

where $m = \gamma - 2$. The relaxation follows a power law dependence with a fractional exponent, $m$. Moreover, $m$ is directly related to the distribution of domain sizes given by the critical exponent, $\gamma$.

### III. EXPERIMENTAL PROCEDURE

To verify the model, the dynamic response of the dielectric susceptibility, polarization, and birefringence of epitaxial $KNbO_3$ was measured. Epitaxial films of $KNbO_3$ for this study were prepared by low-pressure *m*etal-*o*rganic *c*hemical *v*apor *d*eposition (MOCVD) on spinel(100) substrates. The growth procedure and structural characterization have been detailed elsewhere.[27,28] The films are epitaxial as determined by x-ray diffraction and transmission electron microscopy; however, they are multi-domain.[28,29] Film thicknesses ranged from 100 to 400 nm; however, no dependence of the measured transient properties on film thickness was observed. The index of refraction ranged from 2.26 to 2.30, which is comparable to the bulk.[30]

For EO and dielectric measurements, the samples were patterned with Au/Cr, coplanar surface-electrodes separated by a 5 μm gap. The EO effect was measured as a He-Ne laser beam passed through the electrode gap, normal to the film surface. The apparatus used for determination of the EO coefficient and measurement of the transient response has been described thoroughly elsewhere.[3,7] Polarization measurements were carried out on a Radiant Technologies 6000S FE test system.[3] Dielectric measurements were performed on an HP4192A impedance analyzer.[8]

## IV. EXPERIMENTAL RESULTS

The time dependent EO measurements are shown in figure 1. The increase in birefringence, $\delta(\Delta n)$, measured during the application of the 1 µs electric field pulse of 4 MV/m results from domain alignment (inset of figure 1). Upon removal of the field, the domains relax back to their initial configuration over times much longer than the original pulse duration. For each of the films measured, the relaxation response followed a power law of the same form as Eq. (12) with an exponent $m_{EO} \approx 0.2$ to 0.3. (The subscript $EO$ defines the exponent for the EO effect.) The observed power law dependence holds for times spanning the entire measurement range of 6 ns to 0.1 ms.

The dynamic response of the polarization, $\delta P$, for these $KNbO_3$ films is shown in fig. 2. The increase in the polarization, during application of the 1 MV/m, 30 ms electric field pulse, is also due to alignment of domains (inset of fig. 2). After the bias pulse is removed the polarization decays according to a power law. This behavior spanned the resolution limit of the measurement system, from ~1 ms to 0.6 s. For each of the films measured, the power law exponent for the polarization, $m_P$, was between 0.2 and 0.001. The wide range of $m_p$ is attributed to its dependence on the applied pulse duration (1 to 100 ms) and magnitude (0.5 to 2 MV/m), as will be discussed below.

To assess the long-time response of the system, the dielectric response, $\delta\varepsilon$, of the films was also measured as a function of time before, during, and after an applied field of 4 MV/m for 100 s (inset, figure 3). The time dependent behavior of the dielectric relaxation is shown in figure 3. The response is similar to that observed by Sommer and Kleeman in $KTa_xNb_{1-x}O_3$.[6] For times less than 3 s the relaxation follows a power law with $m_\varepsilon \leq 0.1$. For longer times the dielectric response is best described by a modified stretched exponential[15] of the form:

$$\delta \varepsilon(t) = (\varepsilon(0)) t^{-m} \exp\left(-\left(\frac{t}{\tau}\right)^\beta\right) \qquad (13)$$

where $\varepsilon(0)$ is the dielectric constant measured immediately after removal of the bias pulse of 100 s, $\tau$ is the time constant, and $\beta$ is the power of the stretched exponential. The solid line in figure 3 is a fit to this equation with $\tau \approx 96$ s and $\beta \approx 0.34$. The time constant for these films was independent of the applied field duration for pulses from 10 to 1000 s, and $\tau$ ranged from 10 s to 100 s.

## V. ANALYSIS AND DISCUSSION

The results of the transient experiments show a relaxation response consistent with Eq. (11) over a range of times from $10^{-9}$ to ~1 s. However, upon first inspection, the values of the exponents $m_{EO}$ and $m_P$ appear to be distinct. The wide range of measured values (0.3 to 0.001) is actually a result of the different magnitudes and durations of the fields used to align the FE domains. The dependence of $m$ on the applied pulse duration ($m_{EO}$ and $m_P$) and magnitude ($m_P$) is shown in figure 4. This dependence of $m$ on the applied field presumably results from a change in the domain size distribution with applied field strength, $E$, and duration, $t_E$. This suggests that $D_F = D_F(|E|, t_E)$, as observed in relaxor FE systems.[18] For very large $E$ and long $t_E$, a smaller $m$ is measured, which corresponds to the removal of fractal domain features, hence $D_F \rightarrow 3$, $\gamma \rightarrow 2$, and $m \rightarrow 0$. A simple extension of the model presented here would incorporate the dependence on $D_F$ on the applied field.

For the lowest magnitude and shortest duration pulses, the measured $m$ values converge to a value of 0.3 to 0.2, corresponding to the derived values $\gamma$ = 2.3 to 2.2 and $D_F$ = 2.3 to 2.5, respectively. These numbers are well within the framework of the model discussed above, and agree with those predicted by percolation theory, $\gamma$ = 2.2 and $D_F$ = 2.5. Moreover, the derived values are consistent with $D_F$ previously measured in FE[22,23,25] and ferromagnetic films.[31,32] The $m$ values are also comparable to exponents describing the temporal evolution of the domain structure in other FEs.[18,33,34]

The longest time over which the power law relaxation holds, ~1 s, corresponds to the limit imposed by the largest domains. Consistent with Eq. (11), at long times ($t \geq V_{min}/B$) the relaxation is dominated by a stretched exponential transient response, Eq. (13). For measurement times on the order of $t \approx 1$ s the relaxation time of the largest domains, $\tau = V_{max}/B$, is approached. Assuming the largest domain volume in the sample, $V_{max}$, is of the order $(1\ \mu m)^3$, the coefficient $B$ in Eq. (2) must be of the order $1\ \mu m^3 / 1\ s = 10^{-18}\ m^3/s$. Further assuming[13] the volume of the smallest domains, $V_{min}$, is on the order $(1\ nm)^3$ suggests the shortest relaxation times in the system are on the order of $V_{min}/B \approx 1\ nm^3 / 10^{-18}\ m^3/s = 1$ ns. Thus, the predicted relaxation rate of these nano-scale domains corresponds to 1 ns, an observation made by previous investigators.[12,14]

## VI. CONCLUSION

In conclusion, the dynamic response of the EO effect in thin film ferroelectrics has been analyzed for a continuous distribution of FE domain sizes. The analysis indicates the dynamic response is described by the expression $\Phi(t) \propto t^{-m} \exp(-(t/\tau)^{\beta})$ which in the short- time limit reduces to a power law with a fractional exponent. This expression provides an excellent fit for the measured relaxation response of the birefringence, polarization, and dielectric constant for epitaxial $KNbO_3$ thin films on time scales ranging over 11 orders of magnitude, from 6 ns to 1000 s.

**Acknowledgments**

The authors would like to thank Prof. T.J. Marks and Dr. J. Belot for purification of the metal-organic precursor used in film growth.  BMN acknowledges support through the NSF Minority Graduate Fellowship and Lucent Technologies Cooperative Research Fellowship Program.  This work was supported under the MURI program "Integrated Devices for Terabit per Second 1.3 and 1.5 Micron Network Applications" AFOSR/ARPA contract number F49620-96-1-0262 and NSF/MRSEC award number DMR-9632472.

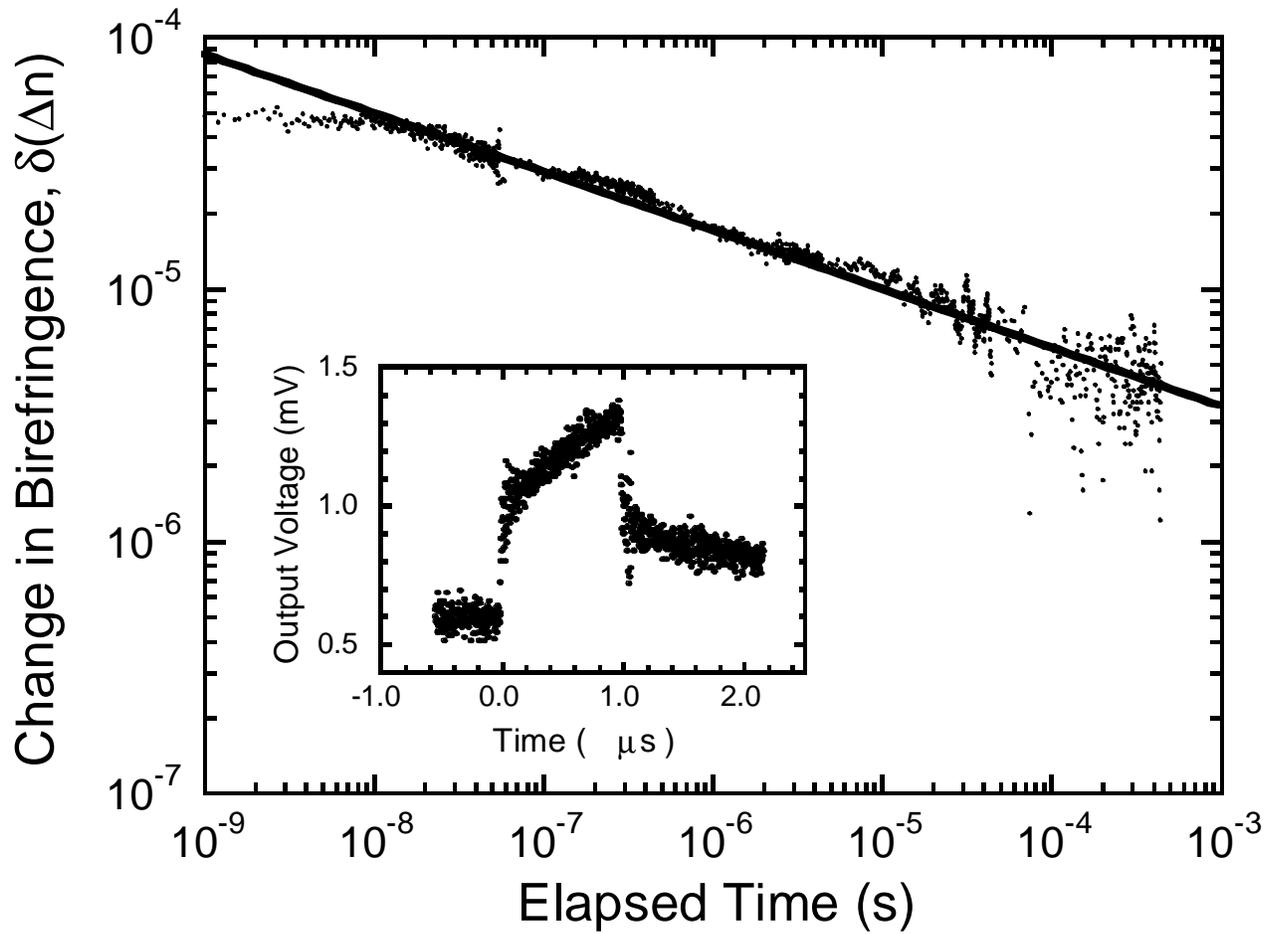

**Figure 1.** The change in birefringence of a $KNbO_3$ thin film after the removal of a 4 MV/m, 1 μs electrical pulse. The data fit a power law (solid line) with an exponent $m_{EO} = 0.2$. The inset shows the measured response upon application and removal of the pulse on a linear scale.

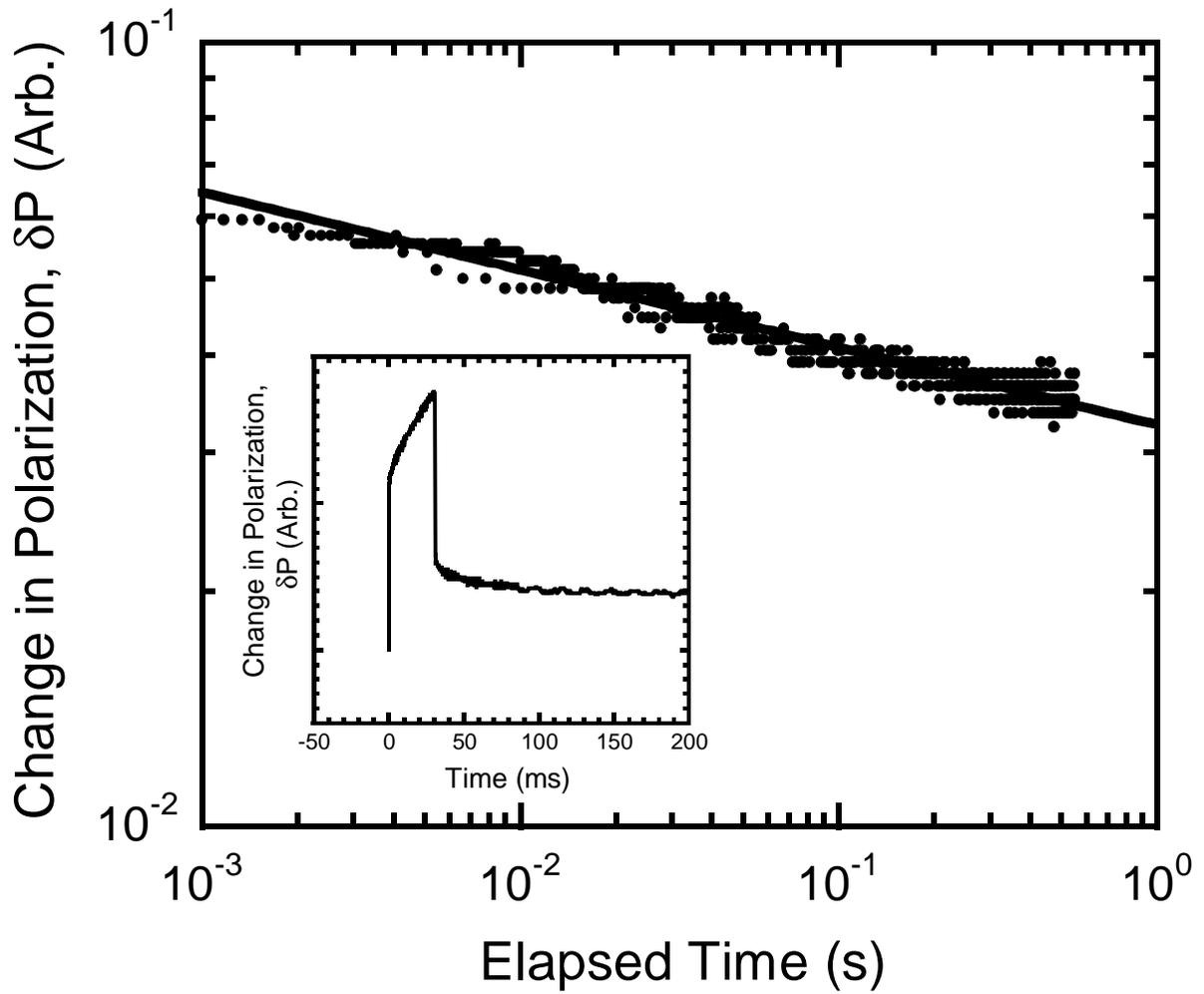

**Figure 2.** The change in the electronic polarization of a KNbO$_3$ thin film after the removal of a 1 MV/m, 30 ms electrical pulse. The data fit a power law (solid line) with an exponent, $m_P = 0.1$. The inset shows the data recorded during application and removal of the pulse on a linear scale.

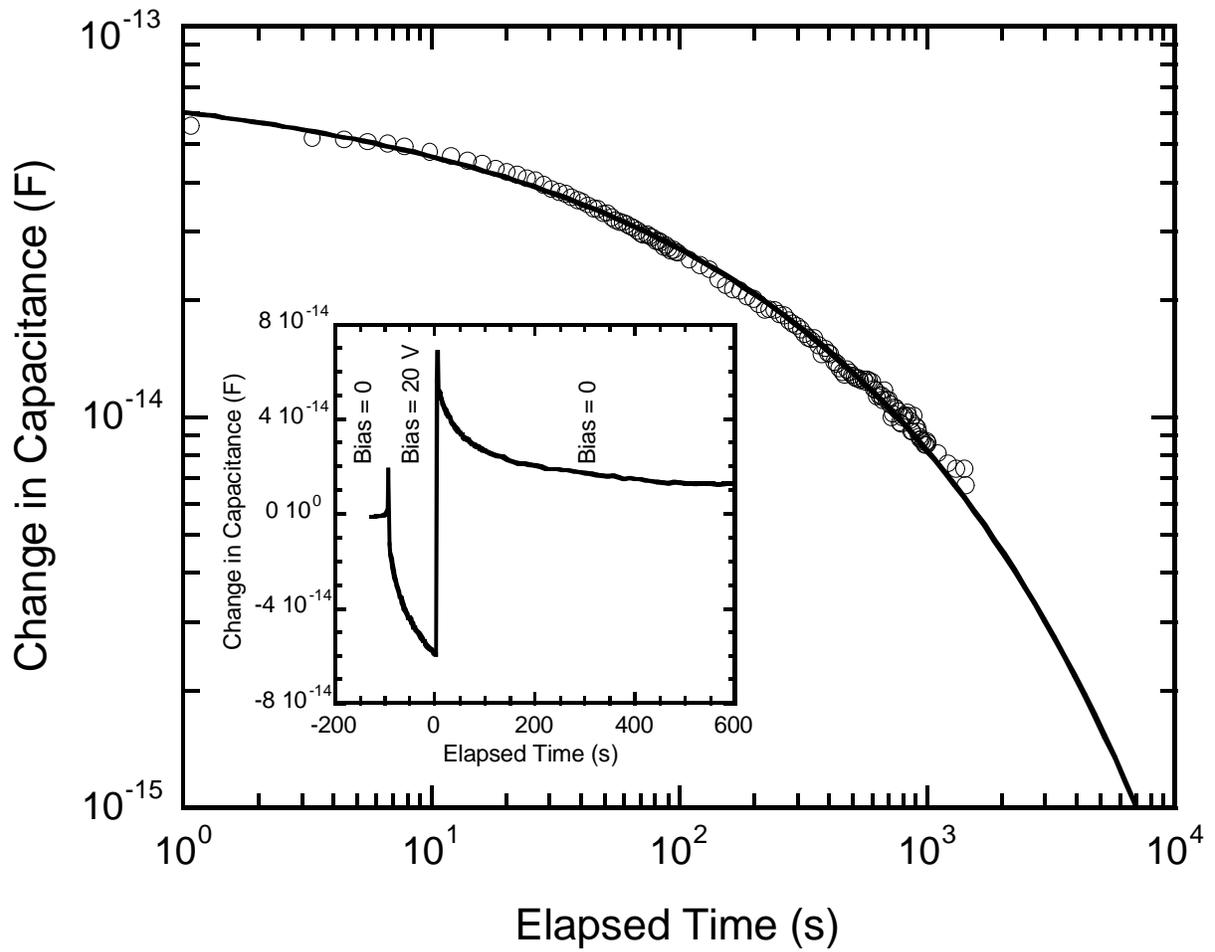

**Figure 3.** The relaxation of the change in capacitance at 1 MHz (1 $V_{pp}$) after removal of the bias field follows a modified stretched exponential as in Eq. 13 (solid line). The inset shows the capacitance before, during, and after application of a 4 MV/m bias for 100 s.

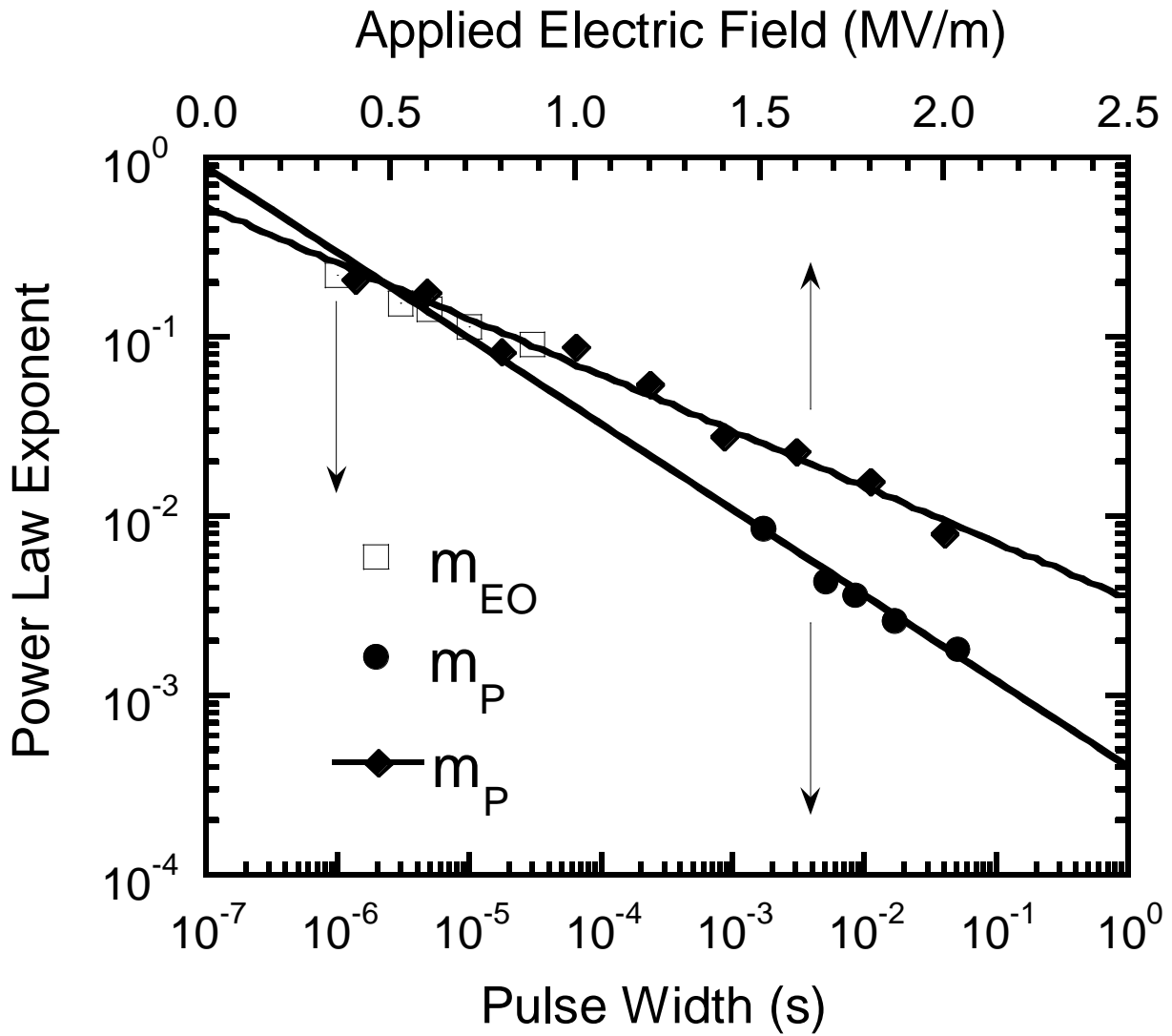

**Figure 4.** Dependence of the exponent $m_{EO}$ (□) on the duration of a 4 MV/m applied pulse and $m_P$ (●) on the duration of a 2 MV/m applied pulse (lower x-axis). Also shown is the dependence of $m_P$ (♦) on the magnitude of a 2 ms applied pulse (upper x-axis). Lines are a guide to the eye.